\documentclass[twocolumn,preprintnumbers,superscriptaddress, amssymb,aps,pra,]{revtex4}
\usepackage{amsmath}
\usepackage{graphicx}
\usepackage{color}
\usepackage[usenames,dvipsnames]{xcolor}
\usepackage{siunitx}
\usepackage[T1]{fontenc}
\usepackage[utf8]{inputenc}
\usepackage{ulem}
\usepackage{subfigure}
\usepackage[colorlinks=true, letterpaper=true, pdfstartview=FitV, linkcolor=red, citecolor=blue, urlcolor=red]{hyperref}

\begin{document}

\title{Chaotic few-body vortex dynamics in rotating Bose--Einstein condensates}

\author{Tiantian Zhang}
\affiliation{Okinawa Institute of Science and Technology Graduate University, Onna-son, Okinawa 904-0495, Japan.}
\author{James Schloss}
\affiliation{Okinawa Institute of Science and Technology Graduate University, Onna-son, Okinawa 904-0495, Japan.}
\author{Andreas Thomasen}
\affiliation{Okinawa Institute of Science and Technology Graduate University, Onna-son, Okinawa 904-0495, Japan.}
\author{Lee James O'Riordan} 
\affiliation{Okinawa Institute of Science and Technology Graduate University, Onna-son, Okinawa 904-0495, Japan.}
\author{Thomas Busch}
\affiliation{Okinawa Institute of Science and Technology Graduate University, Onna-son, Okinawa 904-0495, Japan.}
\author{Angela White}
\affiliation{Okinawa Institute of Science and Technology Graduate University, Onna-son, Okinawa 904-0495, Japan.}
\affiliation{Department of Quantum Science, Research School of Physics and Engineering, The Australian National University, Canberra ACT 2601, Australia.}

\begin{abstract} 
We investigate a small vortex-lattice system of four co-rotating vortices in an atomic Bose--Einstein condensate and find that the vortex dynamics display chaotic behaviour after a system quench introduced by reversing the direction of circulation of a single vortex through a phase-imprinting process.  By tracking the vortex trajectories and Lyapunov exponent, we show the onset of chaotic dynamics is not immediate, but occurs at later times and is accelerated by the close-approach and separation of all vortices in a scattering event.  The techniques we develop could potentially be applied to create locally induced chaotic dynamics in larger lattice systems as a stepping stone to study the role of chaotic events in turbulent vortex dynamics. 
\end{abstract}
\maketitle

\section{Introduction}
Dynamical systems that are highly sensitive to their initial conditions are often referred to as chaotic. Consequently, some of the hallmark indicators of chaotic evolution are a rapid divergence between two trajectories given a small change in initial conditions and a mixing of trajectories in phase space \cite{Strogatz}. Many characteristics of chaos are exhibited in turbulent flows \cite{Spiegel}, including, for example, the exponential separation of pairs of particles in homogeneous isotropic turbulence \cite{Biferale2005}. It has recently been shown that in classical forced turbulence, the degree of chaos depends only on the Reynolds number \cite{Berera2018}.  The importance of chaotic events in classical turbulence and the similarities that can be drawn between classical and quantum turbulence strongly indicate that chaotic events will also contribute to turbulent quantum vortex dynamics, however the nature of these events remains largely unexplored \cite{WhitePNAS}. 

Understanding the chaotic nature of vortex dynamics is therefore an important aspect of developing a detailed understanding of vortex turbulence. At the same time, it also remains an inherently interesting problem in systems composed of such small numbers of vortices that turbulence is absent.  Chaotic few-vortex systems have been known to display intriguing dynamics since the seminal work of Aref and Pomphrey \cite{Aref1980,Aref1982,Aref1983,Aref1988}, who showed that although the two-dimensional hydrodynamics of the classical three vortex system is integrable, this is not true for the case of four identical vortices which can exhibit chaotic dynamics \cite{Aref1982}. The simple nature of quantum vortices in comparison to their classical counterparts has also led to interest in the chaotic nature of quantum vortices in superfluids, such as Bose--Einstein condensates (BECs) and superfluid He \cite{Nemirovskii1995,Hall2013,Kevrekidis2014pre,Kevrekidis2014chaos}. Unlike in the classical case, the onset of chaos in quantum vortices appears with fewer vortices present, and early experiments have explored co-rotating few-vortex clusters in trapped BECs, both for the integrable two-vortex problem as well as the non-integrable three and four vortex cases where chaotic trajectories can occur \cite{Hall2013}. In trapped two-dimensional BECs, the crossover from regular to chaotic motion in a system of three interacting vortices, with two co-rotating vortices and an anti-vortex rotating in the opposite direction has also been explored theoretically through a reduced Hamiltonian approach \cite{Kevrekidis2014pre,Kevrekidis2014chaos}. 

Vortex dynamics in superfluid systems draw remarkable analogies to classical point-vortex systems and classical vortex turbulence in spite of the striking differences between quantum and classical fluids. In BECs, the superfluid velocity, ${\bf v}=\hbar \nabla \theta / m$, is dependent purely on the gradient of the condensate phase $\theta$, which means quantum vortices possess a well-defined strength and quantized circulation, making them simpler than their classical counterparts which can be of any strength and arbitrary circulation. Although quantum and classical vortices are very different, hallmarks of classical turbulence, such as the Kolmogorov spectrum, have been shown to emerge in quantum vortex turbulence on scales larger than the average inter-vortex spacing \cite{Nore1997,Stalp1999,Araki2002,Salort2010}. Vortex turbulence in two-dimensional BECs  has become a subject of increasing experimental interest, due to the controllable nature of BECs and quantum vortices.  Experimental techniques have advanced,  resulting in accessible methods to detect  vortex circulation \cite{Shin2017}, image vortices in-situ \cite{Wilson2015}, and build a picture of vortex dynamics at different times in a single experiment \cite{Freilich2010,Ferrari2017}.  Recent times have seen experiments exhibiting the von K{\'a}rm{\'a}n vortex street \cite{Shin2016}, as well as investigations into decaying two dimensional turbulence \cite{Neely2013,Shin2014} and demonstration of Onsager vortex clusters \cite{Neely2018,Helmerson2018} in Bose--Einstein condensates.  

In this work we will exploit the finely controllable nature of vortices in BECs to extend the picture of chaotic vortex dynamics in systems consisting of small numbers of quantum vortices. Vortex lattices in rotating BECs are extremely robust and have proven to be stable against externally applied perturbations, such as periodic kicking \cite{ORiordan16_moire}.  One way to capitalise on this stability and generate repeatable and well controlled initial conditions is to engineer lattice defects by applying the technique of phase imprinting. Phase imprinting can be used to  control both the number and location of defects in vortex lattices \cite{ORiordan16} and is also experimentally feasible \cite{Dobrek1999}.  Phase imprinting has been applied in a variety of BEC experiments, ranging from transferring orbital angular momentum to atoms from a Laguerre--Gaussian beam  \cite{Andersen2006,Ryu2007}, to using a pulsed light shift with a tailored intensity pattern to create a soliton \cite{Denschlag2000}, vortex \cite{Sengstock1999}, or states with quantized circulation \cite{Perrin2018}. 

In the following, we show that phase imprinting one vortex defect in a small vortex lattice system can be used to controllably  create chaotic vortex dynamics in an experimentally accessible way. We also find that even though chaotic dynamics is confirmed by a positive Lyapunov exponent immediately after the quench, other signals of chaotic dynamics are boosted only at later times by the close approach and consequent scattering of all vortices. Applying this technique to more than one vortex in the lattice may potentially be used to induce chaotic vortex dynamics in systems of larger numbers of vortices, enabling one to control the number of vortices with chaotic trajectories and the location of locally induced vortex chaos in larger lattice systems.  

This manuscript is organized as follows. In Section~\ref{sec:BEC}, we discuss the theoretical model we apply to study the small vortex lattice systems in BECs, introducing the Gross--Pitaevskii equation and describing the technique of phase imprinting. Section~\ref{sec:phase} presents the resulting condensate dynamics after phase imprinting, and discusses how this leads to regular and chaotic vortex dynamics in systems of four co-rotating vortices and three co-rotating vortices and one anti-vortex, respectively. In Section~\ref{sec:measure}, by tracking vortex trajectories and calculating the Lyapunov spectrum, we further quantify the onset of chaos. Finally, our conclusions are presented in Section~\ref{sec:conclusions}.

\section{Theoretical model}
\label{sec:BEC}
The dynamics of vortices in a BEC near zero Kelvin are well described in the mean-field regime by a two-dimensional Gross--Pitaevskii equation (GPE), which can be written as 
\begin{equation} \label{GPE}
i\hbar \frac{\partial}{\partial t} \Psi   = \left( -\frac{ \hbar^2 }{ 2m } \nabla^{2}+\frac{1}{2}m\omega_{\perp}^{2}r^{2}+g |\Psi|^{2} -\Omega L_{z} \right) \Psi  \,,
\end{equation}
in the rotating frame. 
Here $L_{z}=xp_{y}-yp_{x}$ is the angular momentum operator and we choose a realistic value of $\Omega= 0.3\times2\pi $ Hz as the frequency of an externally applied rotation around the $z$--axis.  We model a condensate of  $N=10^{6}$ $^{87}$Rb atoms with an s-wave scattering length $a_s  \approx 90a_0$, confined in an oblate harmonic trapping potential, $(V_{\perp},\ V_z)=\frac{1}{2}m\left(\omega_{\perp}^{2}r^{2},\ \omega_{z}^{2}z^{2}\right)$, with typical trapping frequencies $(\omega_{\perp},\ \omega_{z}) = (2\pi,\ 32\pi)$ Hz and $r^{2}=x^2+y^2$. 
The effective two-dimensional interaction strength is $g=\left(4\pi\hbar^{2}a_{s}N/m\right)\sqrt{m\omega_{z}/2\pi \hbar}=6.8\times10^{-40}\, \text{m}^{4}\text{kg}/\text{s}^2$.

For the evolution of our BEC system, we use the split-step method \cite{Javanainen2006}. In order to find the ground states in the rotating frame, we employ imaginary time propagation by performing a Wick rotation and substituting $t=-i\tau$ in Eq.~\eqref{GPE} \cite{Chiofalo2000}. We then solve this equation efficiently on a spatial grid of $2^{10}\times2^{10}$ points covering an extent of 700$\mu m \times 700\mu m$,
using Graphics Processing Units (GPUs) by making use of the open-source GPE solver, GPUE \cite{GPUE}. 

Under large, externally imposed rotation, the ground-state of a BEC can consist of a triangular lattice of quantum vortices, with the number of constituent vortices dependent on the rotation strength \cite{ChevyDalibard2000,MadisonDalibard2000,Ketterle2001}. However, for smaller rotation strengths other configurations are known \cite{Aftalion2001} and we will concentrate on the regime where four vortices appear in the ground state. While the direction of rotation of the individual vortices is usually the same as that of the applied external rotation, this can be changed by phase imprinting \cite{ORiordan16}.

\begin{figure}[t!]
\includegraphics[width=\columnwidth]{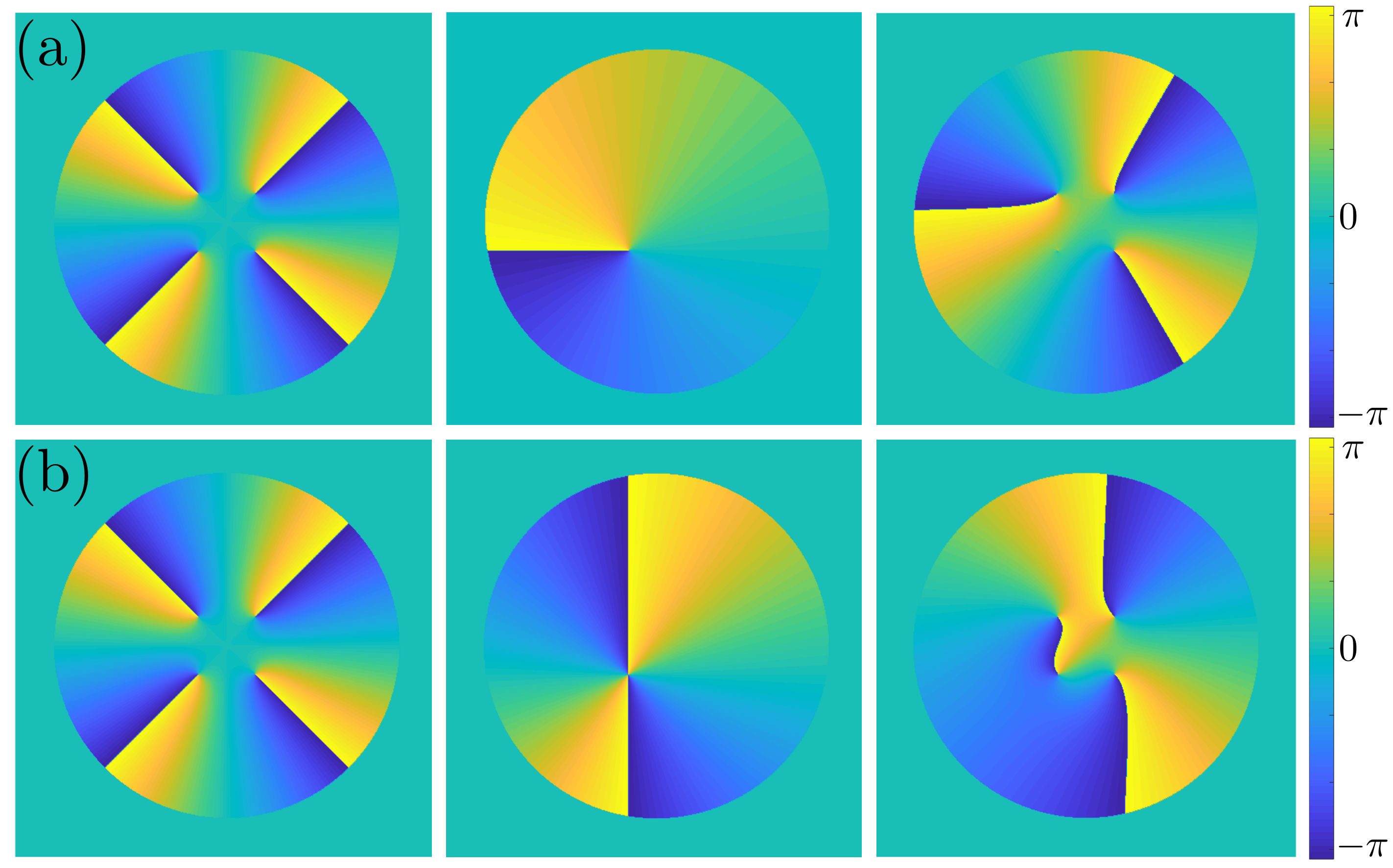}
\caption{From left to right both rows depict the phase profile of the initial co-rotating four-vortex system (left), the phase mask applied to change the circulation of the lower-left vortex (centre) and the resulting four vortex system after the phase imprinting process. (a) Imprinting of a $2\pi$ anti-vortex creates a system of three co-rotating vortices and removes the phase singularity in the lower-left quadrant. (b) Imprinting of a $4\pi$ anti-vortex creates a system of three co-rotating vortices and an anti-vortex in the lower-left quadrant. 
}
\label{Fig:Schematic}
\end{figure}

\begin{figure*}[t!]
  \includegraphics[width=2\columnwidth]{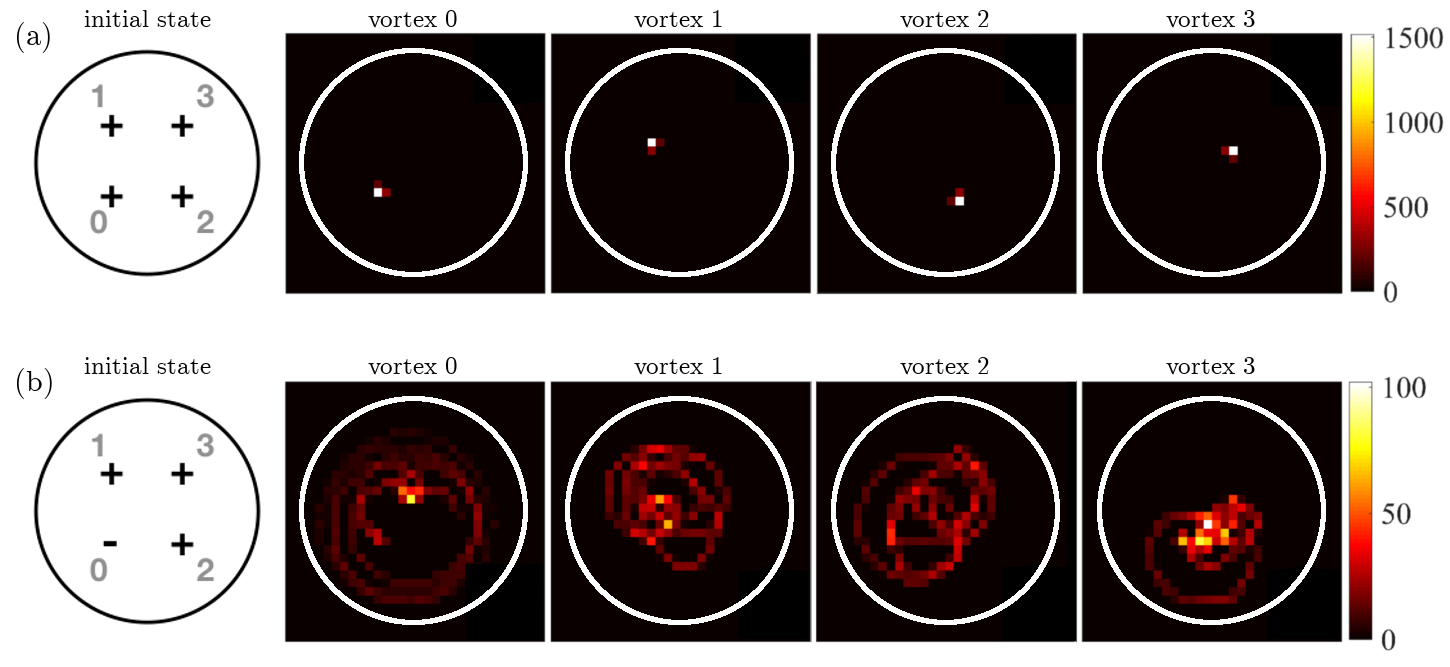}
  \caption{Histograms of the positions of each vortex in the x-y plane, tracked over an evolution time of $20$ seconds in the co-rotating frame. 
The vortex in the lower left has been annihilated by phase imprinting and re-imprinted with (a) the same and (b) the opposite direction of rotation, exactly on the vortex core. The area of each plot is $400\mu m\times 400 \mu m $, and the white circles are contour lines at 40\% of the maximum density, indicating extent of vortex motion for a rotationally symmetric condensate. (a) For four vortices with the same circulation direction, regular trajectories at constant radius appear, while in (b), disordered vortex trajectories can be seen.}
\label{Fig:Trajectories} 
\end{figure*}

Through phase imprinting, a phase profile corresponding to a vortex with $2\pi$ phase winding can be introduced into the condensate by $\Psi_{\text{IMP}}(\textbf{r},t)=|\Psi(\textbf{r},t)|\exp^{i \left(\theta(\textbf{r},t)+ \theta_{\text{IMP}}(\textbf{r})\right)}$ where the four-quadrant $\arctan$ defines $ \theta_{\text{IMP}}(\textbf{r})=\arctan(y-y_{0},x-x_{0})$, which centres the singularity of the vortex phase profile at the position $(x_{0},y_{0})$ in the condensate.  Imprinting a phase mask of $2\pi$ winding centered on an existing vortex of $2\pi$ winding in the same direction will create a doubly charged vortex (of $4\pi$ winding) at that location, which will subsequently decay into two singly charged vortices \cite{Shin2004}. Imprinting a vortex-centred phase mask of $2\pi$ winding in the opposite direction to the vortex circulation will result in vortex annihilation (see Fig.~\ref{Fig:Schematic}(a)), removing the circulation in that region. Finally, imprinting a phase mask with $4\pi$ winding in the opposite direction to the direction of circulation of a vortex on that vortex centre  (see Fig.~\ref{Fig:Schematic}(b)) will reverse the direction of circulation of that vortex. To annihilate or change the direction of circulation of a vortex through phase imprinting, the imprinted phase mask must be centred on a vortex in the lattice close to the vortex core at a distance of less than twice the condensate healing length \cite{ORiordan16}. 

\section{Regular and irregular vortex dynamics}
\label{sec:phase}

\begin{figure}[tb]
  \includegraphics[width=\columnwidth]{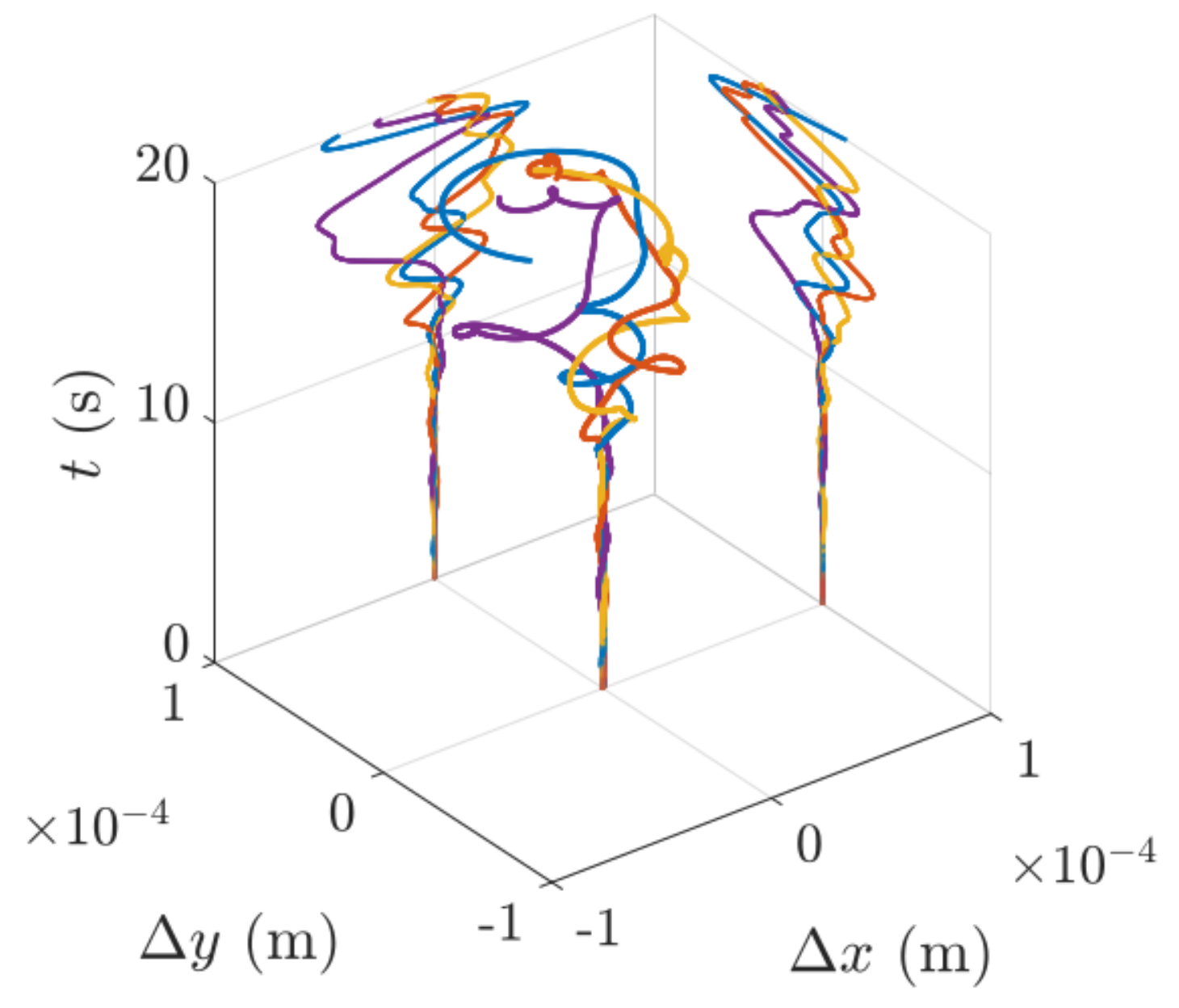}
   \caption{Evolution of the difference in trajectories $\triangle r_i = r_{i}-r'_{i}$ with $i\in \{0,\ 1,\ 2,\ 3\}$ labelling each individual vortex for the four-vortex system. A small change in the initial position of the anti-vortex arises from a phase-imprint at 
   $(x_{0},y_{0})$ and $(x_{0}-\xi/3,y_{0})$ where $(x_{0},y_{0})$ denotes the 
   the pre-existing co-rotating vortex core position. 
   The curves show that even though the onset of disorder is immediate, a strong divergence of trajectories is observed at about $t\approx 10 \si{s}$ (see projections onto the $x-t$ and $y-t$ planes).  The difference in trajectory of the anti-vortex, $\Delta r_0$, is shown in blue, while yellow, orange, and purple lines depict the three co-rotating vortices.
   \label{Fig:Diverging}
}
\end{figure}

\begin{figure}[t]
\includegraphics[width = \columnwidth]{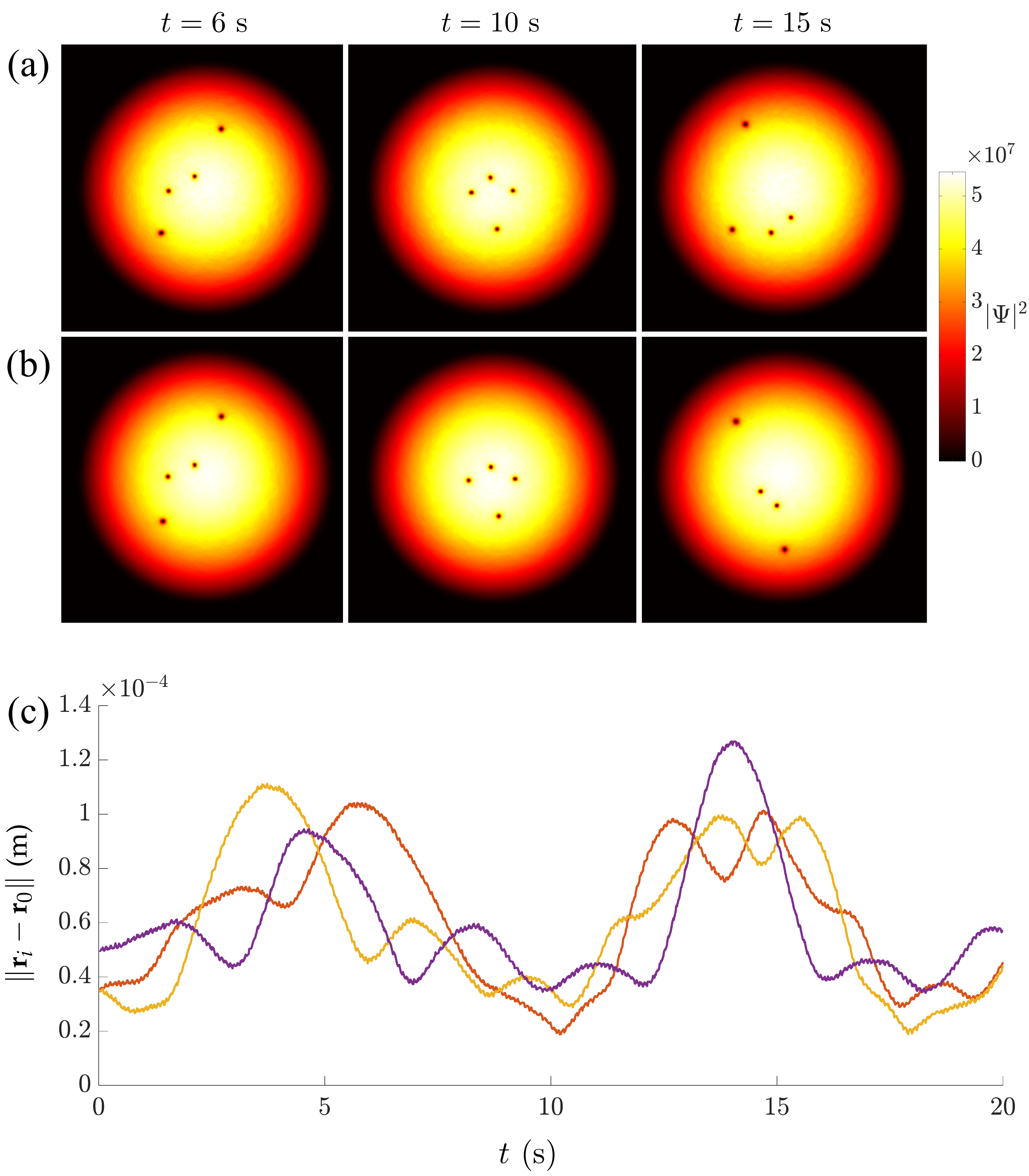}
\caption{Density plots of condensate for (a) $\Delta x = 0$ and (b)$\Delta x=\xi/3$ at times $t=\{6,10,15\}$ s. The densities before the scattering event differ only on small scales (see $t=6s$), whereas for times after the event large deviations are visible (see $t=15s$). At $t=10s$ the vortices make their closest approach. The area plotted is $500 \mu m\times 500 \mu m$. (c) Distances between the vortices at positions $\mathbf{r}_i$ with $i\in \{1,2,3\}$ and anti-vortex at $\mathbf{r}_0$ for $\Delta x = 0$. A minimum around $t=10s$ is clearly visible.
}
\label{Fig:Density}
\end{figure}

\begin{figure}[t!]
\includegraphics[width = \columnwidth]{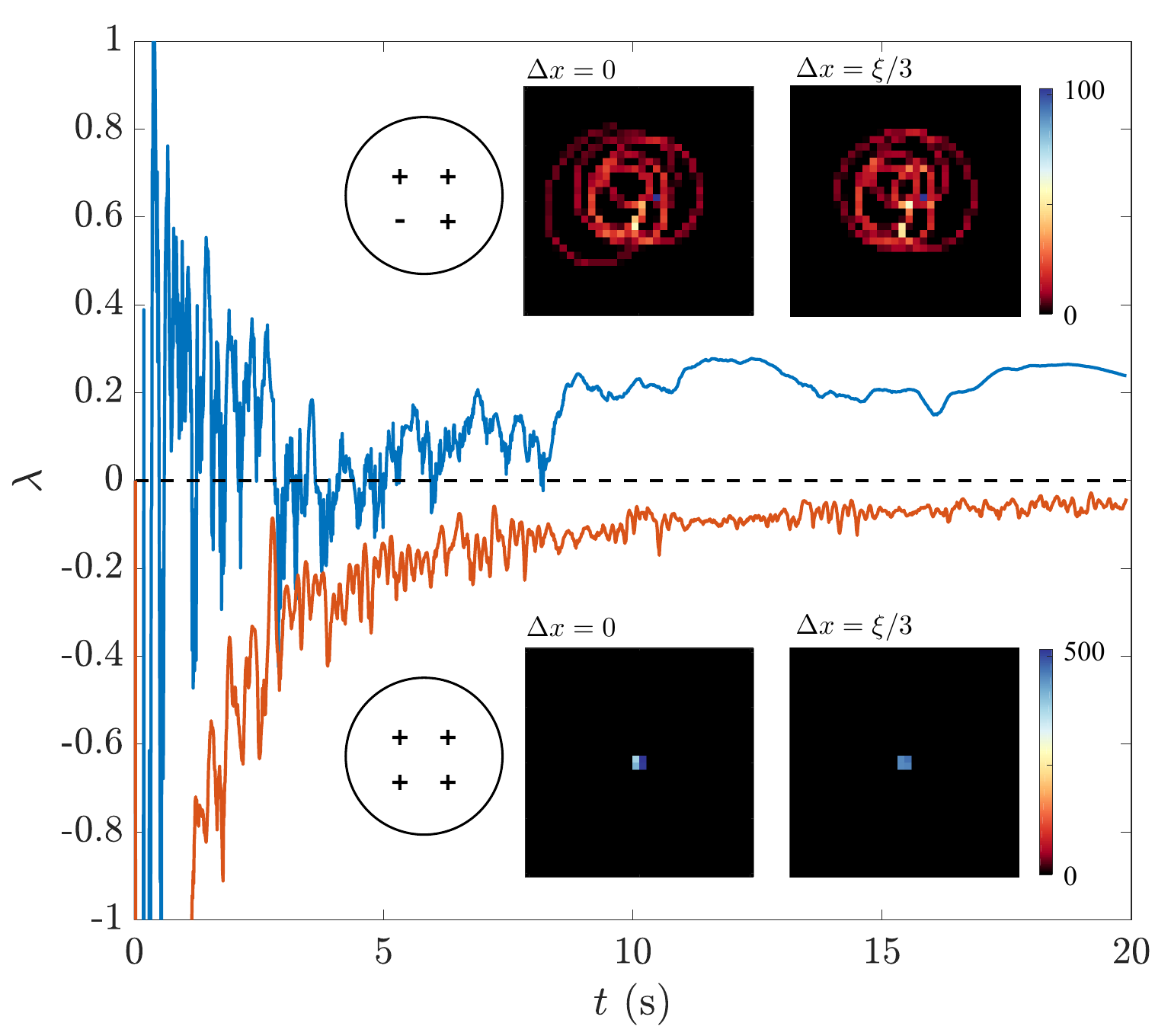}
\caption{The insets show the histograms of the COM trajectories calculated over 20 seconds of evolution for the system of four vortices when the position of a single vortex has been shifted by $\Delta x=0\xi$ and $\Delta x=\xi/3$. The upper two panels depict the corresponding trajectories after the direction of rotation of a single vortex has been reversed, whereas the lower row displays the trajectories for the case where all vortices co-rotate.
The main curve plots the corresponding Lyapunov exponents,  calculated from the shown COM trajectories. 
The negative Lyapunov exponents (orange) indicate that shifting the vortex about the initial position still ensures the stability of vortex trajectories. Reversing the direction of circulation of a single vortex (blue) however leads to fluctuations about zero, eventually leading to a fully positive exponent. 
}
\label{Fig:Lyapunov}
\end{figure}

In this work, we reverse the direction of circulation of a single vortex in a lattice through phase imprinting and investigate if the resulting vortex dynamics are regular or chaotic. To engineer slightly different initial conditions, we vary the distance from the existing vortex core at which the phase imprint has its singularity, while staying within a healing length, and compare the resulting trajectories.  It is well established that a lattice of vortices with the same direction of rotation exhibits regular dynamics \cite{Ketterle2001}, and in Fig.~\ref{Fig:Trajectories}(a) we confirm this for a system of four vortices. The graphs show the histograms of the positions of each individual vortex (labelled 0 to 3 from left to right), where vortex 0 has been erased and re-imprinted with the original winding. One can clearly see that they remain stationary in the co-rotating frame over $20\si{s}$ of evolution, even though a small residual movement from their initial position arises, which stems from the presence of small phonon excitations that were not fully damped out in the numerical ground state finding process. However, when the direction of circulation of a single vortex is reversed in the re-imprinting process, this regularity disappears, and the vortex dynamics become more disordered, with vortices traversing the width of the condensate (shown in Fig.~\ref{Fig:Trajectories}(b)), where the heavily weighted histogram cells indicate more likely vortex positions during evolution. Reversing the direction of rotation of a single vortex can therefore lead to dramatically different results. It is worth noting that such histograms of vortex position evolution could readily be constructed in experiments with available in-situ imaging techniques \cite{Wilson2015,Freilich2010}.   

Although introducing an anti-vortex results in disordered dynamics, one could imagine that this disordered motion is irregular but stable, and any small perturbation in vortex position might give rise to the same irregular trajectories around the condensate. 
To check this we compare two sets of perturbed vortex trajectories with slightly different shifts in the initial position of the anti-vortex, $r_{0}$ and $r'_{0}$ and show the differences in trajectories defined as $\Delta r_i(t) = r_{i}(t)-r'_{i}(t)$ $(i\in \{0,\ 1,\ 2,\ 3\})$ in Fig.~\ref{Fig:Diverging}.
One can see that this difference  is initially small, but starts to diverge significantly at around $t\approx10$s, which is a strong indication of chaotic dynamics in this system.

A close inspection of the vortex dynamics (see supplementary movie \cite{movie}) shows that the strong divergence in trajectories appears to be accelerated when all four vortices move close to each other and are minimally separated.  
Because the velocity fields of each vortex decays as $1/r$, where $r$ is the distance from each vortex core, the vortices experience stronger velocity fields when they are closer to each other; moreover, the point of minimal separation can be interpreted as a multi-vortex {\it scattering} event. Such a scattering event is highly non-linear and accelerates the divergence between the vortex trajectories.  
In Fig.~\ref{Fig:Density} we show snapshots of the full condensate density before ($t=6$s), at ($t=10$s) and after ($t=15$s) the scattering event. One can see that the differences between $\Delta x=0$ (row (a)) and $\Delta x=\xi/3$ (row (b)) are small before, but large after multi-vortex scattering. By tracking the distance between each co-rotating vortex and the anti-vortex, we show in Fig.~\ref{Fig:Density}(c) that a clear minimum around $t=10$s exists, indicating vortex scattering at this time. 
As long as the distance between two vortices is large, the velocity field of one vortex seen by the other vortex will be small, and consequently any small difference stemming from the small change in the initial condition will not have a large effect on the vortex dynamics. 
However, a closer approach of the vortex cores leads to stronger interactions and therefore differences in the velocity fields have a larger influence on the ensuing vortex dynamics, 
leading to the observed boost in the divergence of the resulting trajectories. 
In order to determine if this disordered motion is indeed chaotic, it is necessary to analyse the vortex dynamics in more detail and to do this we proceed to calculate the Lyapunov exponent. 
  
\section{Characterizing chaotic vortex dynamics}
\label{sec:measure}

Lyapunov exponents give the rates of divergence of nearby orbits in phase space \cite{Wolf1985}. 
For the two neighbouring trajectories in four-dimensional phase-space we consider, $\textbf{P}(t)=\left(x(t),y(t),v_{x}(t),v_{y}(t)\right)$ and $\textbf{P}'(t)=\left(x'(t),y'(t),v'_{x}(t),v'_{y}(t)\right)$, with separation defined as $\delta\textbf{P}(t)=\left(\delta x(t),\delta y(t),\delta v_{x}(t),\delta v_{y}(t)\right)$ where $\delta x(t) = x(t)-x'(t)$ ect. The resulting four-dimensional
 Lyapunov exponent, $\lambda$, can then be calculated as
 \begin{equation}
 \lambda=\lim_{t\rightarrow\infty} \frac{1}{t} \ln \frac{||\delta\textbf{P}(t)||}{||\delta\textbf{P}(0)||}, 
 \end{equation}
 where $||\cdot||$ denotes the Euclidean norm. If one or more Lyapunov exponents are positive, the system is chaotic, with unpredictable dynamics developing at a timescale which can be determined from the magnitude of the Lyapunov exponent.  While it is difficult numerically to isolate the evolution of the momentum of each vortex in time, we can build up a representative picture of the phase-space trajectories of individual vortices in our vortex system by tracking both the position and velocity of each vortex.  The Lyapunov exponents can then be calculated by comparing two trajectories with slightly perturbed initial conditions. 

In order to define a phase-space measure for the total system of vortices instead of for each constituent vortex individually, we use a centre of mass (COM) variable defined by $R_{\text{M}} = \frac{1}{n+1}\sum_{i=0}^{n}r_{i}$, where $n+1$ is the number of vortices. Similarly, the centre of mass velocity is expressed as $v_{\text{M}} = \frac{1}{n+1}\sum_{i=0}^{n}v_{i}$.  The insets of Fig.~\ref{Fig:Lyapunov} show the histogram in time (and in the co-rotating frame) of the centre of mass trajectories, which confirm similar irregular dynamics of the system with the anti-vortex (upper row) and regular trajectories of the system with four co-rotating vortices (lower row).
It is worth noting that one could also consider using a global measure which takes into account the direction of circulation of the vortices, and instead consider a {\it centre of charge} variable \cite{Kevrekidis2014chaos}. As we find both quantities give similar qualitative results for calculating the Lyapunov exponents, we only consider the COM of the vortex system here.  

As expected, this COM Lyapunov exponent spectrum, calculated for both regular and disordered four vortex systems, shows that the regular co-rotating system always maintains a negative value for the Lyapunov exponent, indicating non-chaotic behavior (orange curve in Fig.~\ref{Fig:Lyapunov}). However, for the system where a vortex of opposite sign is introduced, we observe that the Lyapunov exponent initially strongly oscillates but later settles and grows to maintain a positive value, indicating chaotic behavior. The time of onset of the constantly positive exponent corresponds to the appearance of the scattering event and the boost in the difference in trajectories. The strong oscillations before that time can therefore most likely be attributed to the finite resolution of our numerics and consequently limited accuracy of the vortex velocities. Such a Lyapunov exponent spectrum could be experimentally constructed through in-situ measurements \cite{Wilson2015,Freilich2010}, where vortex positions (and consequently velocities) are tracked in a single experimental run.

\section{Discussion and conclusions}
\label{sec:conclusions}

We have demonstrated that applying phase imprinting techniques to a small lattice of vortices can change the dynamics of the system from being regular to being chaotic. By simply reversing the direction of rotation of a single vortex, the Lyapunov exponent of the COM trajectories for the vortices clearly indicates the chaotic nature of the vortex trajectories after an initial adjustment phase. In fact, we have found that the chaotic dynamics are accelerated at late times by the close approach and consequent scattering of vortices in a multi-vortex scattering event. This scattering event amplifies the small initial difference in trajectories and results in vortices scattering at different angles while also seeding the ensuing chaotic divergence of trajectories. This method is potentially scalable to larger lattices of vortices and is realisable in cold-atom experiments. 

We have also checked that reversing the direction of circulation of a single vortex in small lattices of five and six vortices still exhibit chaotic dynamics \cite{TiantiansThesis}, and find they exhibit a similar degree of chaotic motion with Lyapunov exponents of 0.24, 0.24 and 0.27 for the four, five and six vortex cases, respectively after 20 seconds of evolution. 
This result contrasts to the limit of rapidly rotating condensates, where large Abrikosov vortex lattices develop and phase-engineering a single vortex defect is known to only induce localised disturbance \cite{ORiordan16}. Studying this crossover from chaotic to regular dynamics, and engineering a crossover from regular to turbulent dynamics is an exciting extension we leave to future work. 

 {\bf{Acknowledgements}:}
This work has been supported by the Okinawa Institute of Science and Technology Graduate University and by JSPS KAKENHI-16K05461 and JP17J01488. We also thank Carlo F.~Barenghi and Makoto Tsubota for useful discussions.

\bibliographystyle{apsrev4-1}

\end{document}